\documentstyle[epsf,sprocl]{article}

\def\Journal#1#2#3#4{{#1} {\bf #2}, #3 (#4)}

\def\NPB{{\em Nucl. Phys.} B}
\def\PLB{{\em Phys. Lett.}  B}
\def\PRL{\em Phys. Rev. Lett.}
\def\PRD{{\em Phys. Rev.} D}

\def\ra{\rightarrow}

\def\be{\begin{equation}}
\def\ee{\end{equation}}
\def\bea{\begin{eqnarray}}
\def\eea{\end{eqnarray}}

\def\dd{\partial}
\def\ep{\epsilon}
\def\la{\raise.16ex\hbox{$\langle$} \, }
\def\ra{\, \raise.16ex\hbox{$\rangle$} }
\def\go{\rightarrow}

\begin{document}

\rightline{\small hep-th/9808105}
\rightline{\footnotesize UMN-TH-1715/98, TPI-MINN-98/14}

\vskip .5cm

\title{COMPLEX MONOPOLES AND GRIBOV COPIES\footnote{To appear in the 
Proceedings of the Third Workshop {\it ``Continuous Advances in QCD''},
Minneapolis, 16 - 19 April, 1998.}}

\author{Y.\ HOSOTANI, K.\ SARIRIAN, AND B.\ TEKIN}

\address{School of Physics and Astronomy, University of Minnesota\\ 
Minneapolis, MN 55455, USA}


\maketitle\abstracts{Complex monopole solutions exist in the  three
dimensional  Georgi-Glashow model with the Chern-Simons term.  They
dominate the path integral and  disorder the Higgs vacuum.
Gribov copies of the vacuum and monopole configurations are studied
in detail.
}

\section{Monopole-supercurrent duality}

In the Georgi-Glashow model, $SO(3)$ gauge theory with a triplet Higgs
scalar field $\vec h$, the gauge sysmetry is spontaneously broken to
$U(1)$ by the Higgs mechanism.    In perturbation theory the $U(1)$ gauge
boson remains massless and the Higgs vacuum is ordered with  
nonvanishing $\la \vec h \ra \not= 0$.

In three dimensional spacetime, however, Polyakov \cite{Polyakov} showed
that  due to instantons, or monopoles in the three-dimensional Euclidean
space,  the Higgs vacuum is disordered $\la \vec h \ra =0$.  The $U(1)$
gauge field acquires a finite mass without breaking the $U(1)$ gauge
invariance. A pair of positive and negative electric charges is bound by an
electric flux tube.  Electric charges are linearly confined.

Physical mechanism at work behind this becomes clear by the duality
transformation.  It was shown by one of the authors \cite{Hosotani} that the
Georgi-Glashow model, or more generally, the compact QED$_3$, is dual to
the  Josephson junction system in the superconductivity.  $(E_1, E_2, B)$
in the compact QED$_3$ corresponds to ($B_1, B_2, E_3)$ in the barrier
region of the Josephson junction.  If a  magnetic monopole-antimonopole
pair is inserted in the barrier region, a magnetic flux of the Abrikosov
magnetic vortex would be formed between the poles as supercurrents 
flow through the barrier.  Instantons (monopoles) in the compact 
QED$_3$ are supercurrents flowing through the two-dimensional barrier in
the Josephson junction.  

\section{Chern-Simons term}

In three-dimensional gauge theory the Chern-Simons term can be added 
to the Lagrangian.  The action of the model is
given by $I = I_{YM} + I_{CS} + I_H$ where $I_{YM}$ and $I_H$ are
the standard Yang-Mills  and Higgs parts of the Georgi-Glashow
model.  The Chern-Simons term is given by
\be
I_{CS} = -{i\kappa\over g^2}\int d^3x ~\epsilon^{\mu \nu \lambda}
~ \hbox{tr}\left( A_\mu \dd_\nu A_\lambda +{2\over 3} A_\mu
A_\nu A_\lambda \right)   ~~.
\label{CSterm}
\ee
In the Euclidean space $I_{CS}$ is pure imaginary for real gauge fields.
It is not gauge invariant.  Under $A \go \Omega A \Omega^{-1} + \Omega
d \Omega^{-1}$, the action $I_{CS}$ changes by
\be
\delta I_{CS} =  {i\kappa\over g^2} \int 
   {\rm tr~} d(A \wedge d\Omega^{-1} \Omega)
+ {i\kappa\over 3g^2} \int {\rm tr~} d\Omega^{-1} \Omega
   \wedge d\Omega^{-1} \Omega \wedge d\Omega^{-1} \Omega ~~.
\label{CSchange1}
\ee
On $S^3$ the first term vanishes.  The second term is proportional to
the winding number, which leads to the quantization of the Chern-Simons 
coefficient;$\,$\cite{Deser} $\kappa = (g^2/4\pi) \times \hbox{integer}$.

What happens to the confinement?  In the 
presence of the Chern-Simons term, the $U(1)$ gauge boson acquires a
topological mass proportional to $\kappa$ even in perturbation theory
so that there occurs no issue of the confinement.  Electric charges are
screened.   How about the Higgs vacuum?  Is the vacuum still disordered
such that $\la \vec h \ra =0$?

In disordering the vacuum, monopole configurations play an important
role.  In the literature it has been argued by many authors 
\cite{D'Hoker,Pisarski,Affleck,Fradkin,Lee,Edelstein,Jackiw} that
monopole configurations suddenly become irrelevant once the Chern-Simons
term is added.  If this is the case, the Higgs vacuum would remain
ordered, i.e.\  the v.e.v.\ of the Higgs field is nonvanishing
and is aligned in one direction in the $SO(3)$ space.  It has been argued
that monopole solutions necessarily have infinite action, and for
configurations of finite action their Gribov copies lead to cancellation. 
We show that this is not the case.  There are complex monopole solutions of
finite action, and Gribov copies do not lead to cancellation.\cite{Tekin}

\section{Monopole ansatz}

The most general monopole ansatz is 
\bea
&&h^a(\vec{x})=\hat x^a h(r) \cr
&&A^a_\mu(\vec{x})= {1\over r} \left[ \epsilon_{a\mu
\nu}\hat{x}^\nu(1-\phi_1) 
+ ( \delta_{a\mu} - \hat{x}_a  \hat{x}_\mu) \phi_2 
    + rS \hat{x}_a  \hat{x}_\mu\right]       
\label{configuration1}       
\eea 
where $\hat x^a = x^a/r$.
Field strengths are
\bea
F_{\mu\nu}^a &=&
 {1\over r^2}
  \ep_{\mu\nu b}  \hat x^a \hat x^b (\phi_1^2 + \phi_2^2 -1)
+ {1\over r} (\ep_{a\mu\nu} -  \ep_{\mu\nu b}  \hat x^a \hat x^b )
(\phi_1' + S\phi_2) \cr
\noalign{\kern 10pt}
&& \hskip 4cm + {1\over r}  (\delta^{a\nu} \hat x^\mu
 -\delta^{a\mu} \hat x^\nu)  (\phi_2' - S\phi_1) ~.
\label{FieldStrength1}
\eea

There is residual $U(1)$ gauge invariance.  Under
$\Omega=\exp \big\{ {i\over 2} f(r) \hat x^a \sigma^a \big\} $
\be
\pmatrix{\phi_1\cr \phi_2\cr} \go 
\pmatrix{\cos f & \sin f \cr -\sin f & \cos f \cr} 
  \pmatrix{\phi_1\cr \phi_2\cr} ~~,~~
S ~~ \go ~~ S - f' ~~.
\label{transformation1}
\ee
The Chern-Simons term transforms as
\be
I_{CS} \go I_{CS} + {4\pi i\kappa\over g^2} \big\{ f(\infty)  - f(0)
\big\} ~.
\ee
Hence gauge copies of solutions carry the oscillatory factor in
$e^{-I_{CS}}$ in the path integral, which
could lead to the cancellation among monopole contributions.

However, the gauge is fixed in the path integral.  For instance,
in the radial gauge $\hat x^\mu A^a_\mu = \hat x^a S =0$, there remains
no residual gauge freedom, once the boundary condition $f(0)=0$ is imposed
for the regularity of configurations.  The radiation gauge is more subtle, 
and is discussed below in more detail.

The action before the gauge fixing is 
\bea
I = {4\pi\over g^2} \int_0^\infty dr \, \Big\{(\phi_1' + S\phi_2)^2 +
(\phi_2' - S\phi_1)^2 + {1\over 2r^2} (\phi_1^2 + \phi_2^2 -1)^2 &&\cr
\noalign{\kern 5pt}
+i \kappa \Big[ \phi_1' \phi_2 - \phi_2'(\phi_1 -1) + S(\phi_1^2 +
\phi_2^2 -1) \Big] &&\cr
\noalign{\kern 5pt}
+{r^2\over 2} h'^2 + h^2 (\phi_1^2+\phi_2^2) + {\lambda r^2\over 4}
(h^2-v^2)^2 \Big\} &.&
\label{action1}
\eea

\section{Complex monopole solution}

The action (\ref{action1}) is a functional of four functions
$\phi_1, \phi_2, S$ and $h$.    In the path integral the gauge fixing
condition is inserted;
\be
Z = \int {\cal D}A^a_\mu {\cal D}\vec h 
~ \Delta_F[A] ~\delta[F(A)] ~ e^{-I} ~.
\label{PI}
\ee
We look for configurations which extremize the action $I$ within
the subspace specified with $F(A)=0$.  In the monopole ansatz
 one, or one combination, of $\phi_1, \phi_2$ and $ S$ is eliminated
by gauge fixing so that three equations need to be solved.

In the radial gauge $S=0$ the extremization of the action leads to
\bea
&&\phi_1'' + {1\over r^2} (1-\phi_1^2  -\phi_2^2) \phi_1
 + i\kappa \phi_2' - h^2 \phi_1 = 0 \label{EqMotion6} \cr
&&\phi_2'' + {1\over r^2} (1-\phi_1^2  -\phi_2^2) \phi_2
 - i\kappa \phi_1' - h^2 \phi_2 = 0  
\label{EqMotion1}
\eea
and 
\be
{1\over r^2} {d\over dr} \Big( r^2 {dh\over dr}  \Big) 
- \lambda (h^2 - v^2) h -
{2\over r^2} (\phi_1^2 + \phi_2^2) h = 0 ~~.
\label{EqMotion2}
\ee
Since eq.\ (\ref{EqMotion1}) contains complex terms,
solutions necessarily become complex. The boundary conditions
ensuring the regularity of configurations and the finiteness of the action
are $\phi_1(0)=1$, $\phi_2(0)=h(0)=0$, $\phi_1(\infty)=\phi_2(\infty)=0$,
and $h(\infty)=v$.

\begin{figure}[th]\centering
\mbox{
\epsfysize=3in \epsfbox{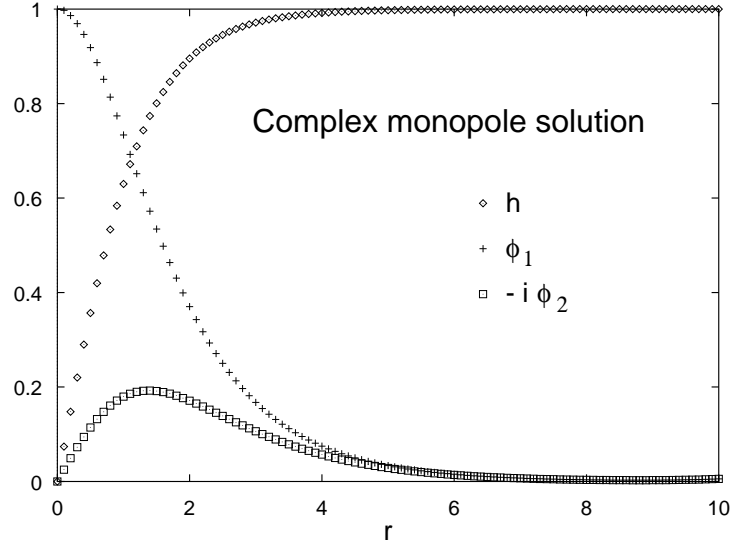}}
\caption{Complex monopole solution in the radial gauge for $v=1$, $\kappa=.5$, 
and $\lambda=.5$.  $\phi_2(r)$ is pure imaginary.} 
\end{figure}

Eqs.\ (\ref{EqMotion1}) and (\ref{EqMotion2}) can be solved by 
an ansatz
\be
\phi_1 = \zeta(r) \cosh {\kappa r\over 2} ~~,~~
\phi_2 = i \zeta(r) \sinh {\kappa r\over 2} ~~.
\label{ansatz1}
\ee
The equations to be solved are
\be
\zeta'' - {1\over r^2} (\zeta^2 -1) \zeta 
   - \Big( h^2 + {\kappa^2\over 4} \Big) \zeta = 0
\label{EqMotion3}
\ee
and eq.\ (\ref{EqMotion2}) where $\phi_1^2 + \phi_2^2=\zeta^2$.
The solution to these equation can be found numerically and
depicted in  fig.\ 1.  $\phi_2(r)$ is pure imaginary.  The action 
for this configuration is real and finite.  The $U(1)$ field
strengths are given by $F^{U(1)}_{\mu\nu}= - \ep_{\mu\nu\rho} \hat
x^\rho/r^2$, exactly those of a magnetic monopole.  Non-Abelian field
strengths are complex.  We call it a complex monopole configuration.

What is the significance of complex monopoles?  In the original form 
of the path integral, field configurations to be integrated are real.
It is an infinite dimensional integral defined along real axes.
Symbolically  we have an integration $\int_{-\infty}^\infty dz \,
e^{-f(z)}$ where
$z$ stands for a field variable, say, $A_\mu(x)$.  The saddle 
point, $z_0$, of $f(z)$ may be located off the real axis.  In the 
saddle point method for the integration, the integration path is
deformed such that a new path pass $z_0$.  $e^{-f(z_0)}$ gives
a dominat factor, approximating the integral.  Translated
in the field theory model, the complex monopole configuration
approximates the integral.  

Complex monopole configurations dominate the path integral. 
They are relevant in disordering the Higgs vacuum.  Without
monopole-type configurations the perturbative Higgs vacuum cannot
be disordered and $\la \vec h \ra$ remains nonvanishing.  With
complex monopoles taken into account $\la \vec h \ra =0$ but $\la
\vec h^2 \ra \sim v^2$.

\section{Gribov copies}

The radiation gauge does not uniquely fix gauge field
configurations.\cite{Gribov}  Even in the monopole ansatz
(\ref{configuration1}), there remains arbitrariness.   The radiation gauge
condition $\dd_\mu A^a_\mu =0$ is satisfied if $ (r^2 S)' = 2\phi_2$. 
Under $U(1)$ gauge transformation (\ref{transformation1}), the radiation
gauge condition  is maintained, provided $f(r)$ obeys
\be
f'' + {2\over r} f' - {2\over r^2} \Big\{ \phi_1 \sin f
+ \phi_2 (1-\cos f) \Big\} = 0 ~~.
\label{Gribov1}
\ee
 $f(0)$ is assumed to vanish for the regularity. Solutions to eq.\
(\ref{Gribov1}) define Gribov copies.   $ A^f = \Omega A \Omega^{-1} +
\Omega d\Omega^{-1}$ is on a gauge orbit of $A$ within the radiation gauge
slice.

These copies have a significant effect in the Chern-Simons theory.
The Chern-Simons term is not gauge invariant as displayed
in (\ref{CSchange1}). We suppose that the Chern-Simons
coefficient is quantized; $4\pi \kappa/g^2=n$ is an integer.
Each Gribov copy carries an extra factor $e^{-\delta I_{CS}}=
e^{-inf(\infty)}$ in the path integral.  Hence, if $f(\infty)$ takes
continuous values, then Gribov copies of monopole configurations
may lead to the cancellation in the path integral.   Indeed,
the authors of ref.\ 6 and ref.\ 7 have
argued that because of this effect contributions of monopole
configurations disappear once the Chern-Simons term is added.
One has  to examine Gribov copies more carefully.

\begin{figure}[ht]\centering
\mbox{
\epsfysize=3.2in \epsfbox{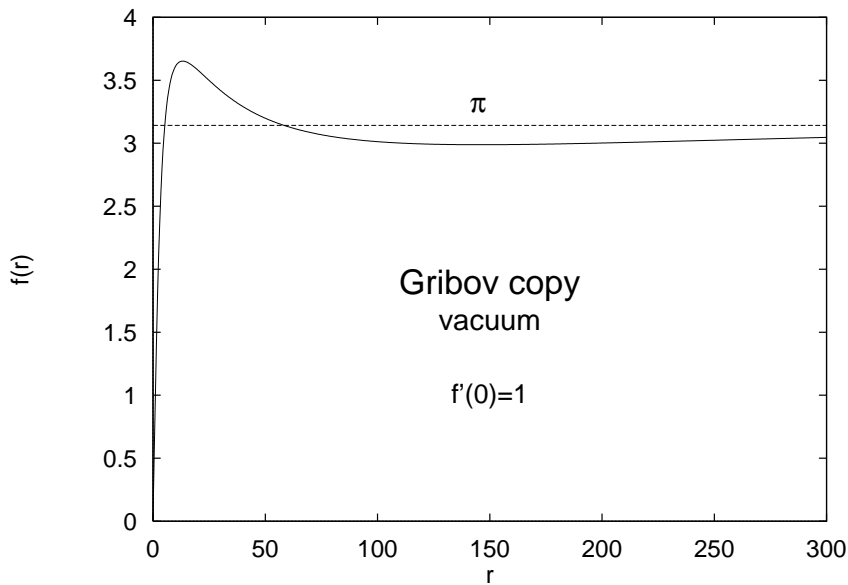}}
\caption{Gribov copies of the vacuum.  The solution $f(r)$ to eq.\ (14)
with $f'(0)=1$ is depicted.  It asymptotically approaches $\pi$.} 
\end{figure}

\subsection{Vacuum}

Before discussing Gribov copies of monopoles, it is worthwhile to
recall Gribov copies of the vacuum.  $A^a_\mu=0$ corresponds
to $\phi_1=1$ and $\phi_2=S=0$.  Eq.\ (\ref{Gribov1}) reads 
\be
f'' + {2\over r} f' - {2\over r^2}  \sin f =0 ~~.
\label{Gribov2}
\ee
The equation is scale invariant.  If $f(r)$ is a solution,
then $g(r)=f(ar)$ also solves the equation.   Solutions are
parametrized by $f'(0)$. Given the initial condition $f(0)=0$ and
$f'(0)$, the solution is uniquely determined.

In fig.\ 2 the solution is depicted for $f'(0)=1$.  For $f'(0)
\equiv \alpha >0$, $f(r)$ reaches the maximum value $3.652$ 
 at $\alpha r=13.2$, then decrease to a local minimum $2.988$
at $\alpha r=145.9 $, and then asymptotically approaches $\pi$. 
For $f'(0) <0$,  $f(\infty)=-\pi$.

\begin{figure}[ht]\centering
\mbox{
\epsfysize=3.2in \epsfbox{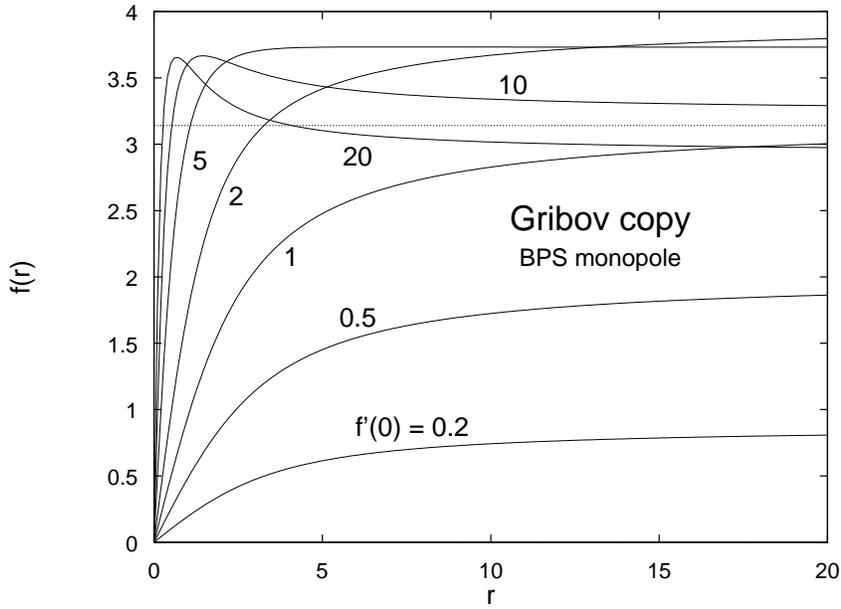}}
\caption{Gribov copies of the BPS monopole.  Solutions $f(r)$ to
eq.\ (13) with the BPS monopole configuration $\phi_1$ with various
values of $f'(0)$.} 
\end{figure}

\subsection{Monopoles}

Consider a monopole in the BPS limit ($\lambda=\kappa=0$) in which
$\phi_1=vr/\sinh vr$.  In this case $\phi_1$ dies out quickly for
large $vr$.  The asymptotic value $f(\infty)$ depends on the
initial slope $f'(0)$.  Solutions $f(r)$ are depicted in fig.\ 3.

In fig.\ 4 $f(\infty)$ is plotted as a function of $f'(0)$. The 
range of the asymptotic value is $-3.98 < f(\infty) < + 3.98$.
For large $|f'(0)|$, $f(\infty) = \pm \pi$.  It is quite unlikely
that, in the presence of the Chern-Simons term, these Gribov copies
of the BPS monopole lead to the cancellation $\sum e^{-inf(\infty)}
=0$. Monopole configurations remain important in the path integral.

\begin{figure}[ht]\centering
\mbox{
\epsfysize=3.2in \epsfbox{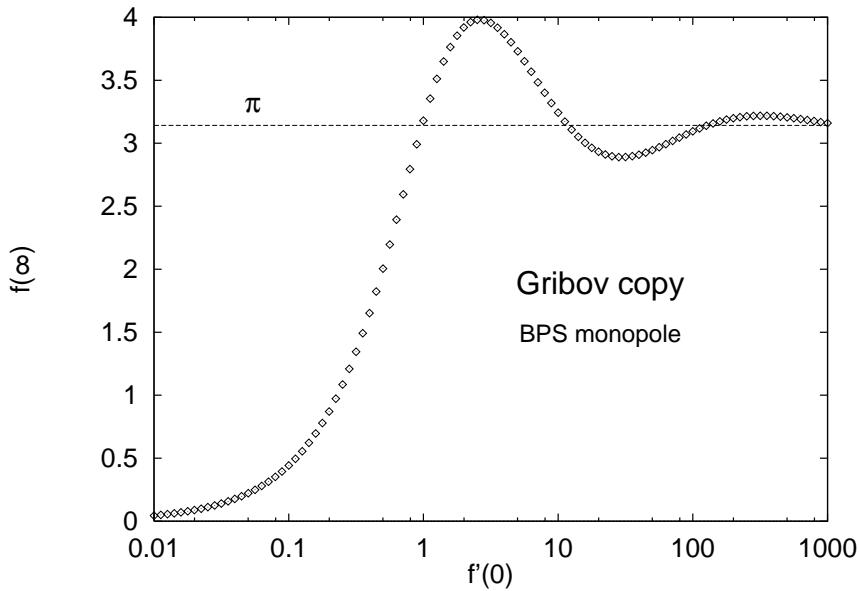}}
\caption{$f'(0)$ vs $f(\infty)$ for Gribov copies of the BPS monopole.} 
\end{figure}

\subsection{Complex monopoles}

There seem complex monopole solutions in the radiation gauge as
well.  Most likely $\phi_2$ and $S$ are pure imaginary.  It is 
legitimate to ask if there are Gribov copies of complex monopoles.

It is obvious that there is no real solution $f$ to
the Gribov equation (\ref{Gribov1}), as $\phi_2$ is pure 
imaginary.  There is no real gauge copy.  

However, this does not necessarily mean that the solution is unique
in the  radiation gauge.  There may be solutions which are related by
``complex'' gauge transformation generated by complex $f(r)$.  

Indeed, one can show that $(A^f, h)$ satifies the equations in the
radiation gauge, and has the same real action as $(A,h)$.
There are many ``complex'' Gribov copies of complex monopoles.
Among them, copies generated by $f$ satisfying Im$\,f(\infty)=0$
may play an important role.
Of course at one or higher loop level, these copies are believed to yield
different contributions in the path integral.   Deeper understanding
is necessary about complex monopoles.

\section{Conclusion}

In the Georgi-Glashow-Chern-Simons model complex monopole
solutions of finite action exist in the radial gauge.  These complex
monopoles dominate in the path integral, and disorder the Higgs
vacuum.

Solutions seem to exist in the radiation gauge as well.  We have
examined Gribov copies of both real and complex monopole
configurations.  The asymptotic value of the gauge function $f(r)$
specifying Gribov copies of monopoles depends on $f'(0)$.    The
Chern-Simons term sensitively depends on $f(\infty)$, but it will
not diminish the relevance of monopole configurations.

It is a fascinating problem to find  an analog of the Chern-Simons term in
the Josephson junction system.   The relevance of monopoles is translated to
the relevance of supercurrents in the Josephson junction system
by the duality transformation.

\section*{Acknowledgments}
The authors would like to thank Bob Pisarski and 
Pierre van Baal for useful discussions in the workshop.  This work was
supported in part  by  the U.S.\ Department of Energy under contracts
DE-FG02-94ER-40823.

\def\AP{{\em Ann.\ Phys.\ (N.Y.)} }


\section*{References}
\begin{thebibliography}{99}

\bibitem{Polyakov}
A.\ M.\ Polyakov, \Journal{\PLB}{59}{80}{1975};
\Journal{\NPB}{120}{429}{1977}. 

\bibitem{Hosotani}
Y.\ Hosotani, \Journal{\PLB}{69}{499}{1977}.

\bibitem{Deser}
S.\ Deser, R.\ Jackiw and S.\ Templeton, \Journal{\PRL}{48}{976}{1982}.

\bibitem{D'Hoker} 
E.\ D.\ D'Hoker and L.\ Vinet, \Journal{\AP}{162}{413}{1985}. 

\bibitem{Pisarski}
R.\ D.\ Pisarski, \Journal{\PRD}{34}{3851}{1986}. 


\bibitem{Affleck}
I.\ Affleck, J.\ Harvey, L.\ Palla and G.\ Semenoff, 
\Journal{\NPB}{328}{575}{1989}.

\bibitem{Fradkin}
E.\ Fradkin and F.\ A.\ Schaposnik, \Journal{\PRL}{66}{276}{1991}.

\bibitem{Lee}
K.\ Lee, \Journal{\NPB}{373}{735}{1992}.

\bibitem{Edelstein} J.\ D.\ Edelstein and F.\ A.\ Schaposnik, 
\Journal{\NPB}{425}{137}{1994}.

\bibitem{Jackiw} 
R.\ Jackiw and S.Y.\  Pi, \Journal{\PLB}{423}{364}{1998}.

\bibitem{Tekin} B.\ Tekin, K.\ Saririan, and Y.\ Hosotani,
preprints hep-th/9808045; hep-th/9808057.

\bibitem{Gribov} V.\ N.\ Gribov, \Journal{\NPB}{139}{1}{1978}.




\end{thebibliography}
\end{document}